\input epsf
\documentstyle[12pt,aasms4]{article}
\doublespace

\newcommand{\beq}{\begin{equation}}
\newcommand{\eeq}{\end{equation}}
\newcommand{\eeql}[1]{\label{eq:#1}\end{equation}}

\newcounter{compteur}

\def\gcc{\rm g\,cm^{-3}}

\def\mv{M_V}
\def\mj{M_J}
\def\mbol{M_{bol}}
\def\814{I_{814}}
\def\606{R_{606}}
\def\555{V_{555}}

\def\msol{M_\odot}
\def\lsol{L_\odot}
\def\llsol{\log (L/L_\odot)}

\def\te{T_{\rm eff}}
\def\simgr{\,\hbox{\hbox{$ > $}\kern -0.8em \lower 1.0ex\hbox{$\sim$}}\,}
\def\simle{\,\hbox{\hbox{$ < $}\kern -0.8em \lower 1.0ex\hbox{$\sim$}}\,}
\def\wig#1{\mathrel{\hbox{\hbox to 0pt{%
          \lower.5ex\hbox{$\sim$}\hss}\raise.4ex\hbox{$#1$}}}}

\begin{document}

\title{{\bf Cooling sequences and color-magnitude diagrams for cool white dwarfs with hydrogen-atmospheres}}

\author{ {\sc G. Chabrier$^{1}$, P. Brassard$^{2}$ , G. Fontaine$^{2}$ and D. Saumon$^{3}$}}

\bigskip
 
$^{1}${Ecole Normale Sup\'erieure de Lyon (C.R.A.L; UMR 5574 CNRS),\\ \indent\indent  69364 Lyon
 Cedex 07, France 

$^{2}$D\'epartement de Physique, Universit\'e de Montr\'eal, C.P. 6128\\
\indent\indent Succ. A., Montr\'eal, Qu\'ebec, Canada H3C 3J7

$^{3}$Department of Physics and Astronomy, Vanderbilt University,\\
\indent\indent P.O. Box 1807 Station B, Nashville, TN 37235}

\begin{center}
$\rm\underline{Accepted\ to}$: {\sl Astrophysical Journal}

\bigskip

\end{center}
\bigskip
\bigskip

\begin{abstract}

We present new cooling sequences, color-magnitude diagrams, and 
color-color diagrams for cool white dwarfs with pure hydrogen 
atmospheres down to an effective temperature $\te=1500$ K. We include a
more detailed treatment of the physics of the fully-ionized interior,
particularly an improved discussion of the thermodynamics of the 
temperature-dependent ion-ion and ion-electron contributions of the 
quantum, relativistic electron-ion plasma. The present calculations 
also incorporate accurate boundary conditions between the degenerate 
core and the outermost layers as well as updated atmosphere models 
including the H$_2$-H$_2$ induced-dipole absorption. We examine the 
differences on the cooling time of the star arising from uncertainties 
in the initial carbon-oxygen profile and the core-envelope $L$-$T_c$ 
relation. The maximum time delay due to crystallization-induced chemical 
fractionation remains substantial, from $\sim 1.0$ Gyr for 0.5 and 1.2 $\msol$ white dwarfs to $\sim 1.5$ Gyr for 0.6 to 0.8 $\msol$ white dwarfs, even with 
initial stratified composition profiles, and cannot be ignored in 
detailed white dwarf cooling calculations. These cooling sequences 
provide theoretical support to search for or identify old disk or halo 
hydrogen-rich white dwarfs by characterizing their mass and age from their 
observational signatures.

Subject headings : equation of state - stars: atmospheres - white dwarfs - Galaxy: halo
\end{abstract}

\section{Introduction}

The possible interpretation of the faint blue objects in the Hubble Deep Field
as halo white dwarfs (Hansen 1998; Ibata et al. 1999; M\'endez \& Minniti 2000), consistent with
the interpretation of the observed microlensing events towards the Large Magellanic Clouds as
stellar remnants (Chabrier, Segretain, \& M\'era 1996; Adams \& Laughlin 1996;
Graff, Laughlin \& Freese 1998; Chabrier 1999), and
the spectroscopic identification of very cool ($\te\simle 4000$ K), high proper motion white dwarfs (WD) (Hodgkin et al. 2000; Ibata et al. 2000)
has triggered interest in the study of old,
cool WDs and stressed the need for
accurate cooling sequences and predicted observational signatures for these objects.
This implies a correct cooling theory for crystallized  white dwarfs
and reliable atmosphere models and
photometric predictions.

The basic physics entering WD evolution has evolved significantly since the pioneering
work of Mestel \& Ruderman (1967) and the
first detailed evolutionary calculations by Lamb \& Van Horn (1975). 
Noteworthy advances have been made on the fronts of
the conductive opacity (Itoh et al. 1983; Itoh, Hatashi, \& Kohyama 1993; Potekhin et al. 1999), the radiative opacity
(Lenzuni, Chernoff, \& Salpeter 1991; Rogers \& Iglesias 1992), the
envelope equation of state (EOS) (Saumon, Chabrier, \& Van Horn 1995, SCVH),
and the detailed description of the thermodynamic properties of the
dense, fully ionized interior plasma, including the main effects of ion
crystallization, namely the latent heat and the chemical fractionation 
(Segretain et al. 1994 and references therein). In the meantime, 
substantial improvement in the theory of the atmosphere of cool WDs 
has been accomplished (Bergeron, Wesemael, \& Fontaine 1991; Bergeron, 
Saumon, \& Wesemael 1995, hereafter BSW; Bergeron, Wesemael, \&
Beauchamp 1995, hereafter BWB). 

These studies were devoted primarily to the characterization of the disk WD population and were
thus restricted to effective temperatures $\te > 4000$ K.
As identified initially in BSW, the onset of molecular recombination below $\sim 5000$ K
and the pending Collision-Induced Absorption (CIA) due to H$_2$-H$_2$ and H$_2$-He collisions
in such dense atmospheres ($g\simeq 10^8$ to $10^9$ cm$\,$s$^{-2}$) 
results in important departures from a blackbody energy distribution, with an increasing absorption of the 
flux longward of 1 $\mu$m.
The WD atmosphere calculations were extended recently to lower temperatures
by Hansen (1998, 1999) and by Saumon \& Jacobson (1999, hereafter SJ), reaching the effective temperature 
range characteristic of the halo
or globular cluster old WD population. As shown by these authors, the strong CIA in the infrared
redistributes the flux toward shorter wavelengths
so that the emergent flux peaks in optical passbands, regardless of $\te$.
This effect has been confirmed recently with the spectroscopic observation of  H$_2$ CIA in LHS 1126 (Bergeron, Ruiz \& Leggett, 1997), LHS 3250 (Harris et al. 1999) and WD0346+246 (Hodgkin et al. 2000). 
As demonstrated initially by Hansen (1998), the colors of very cool WDs are consistent
with the unidentified faint blue objects in the Hubble Deep Field (HDF) (Bahcall et al. 1994; 
M\'endez et al. 1996; Nelson et al. 1996; Ibata et al. 1999). A preliminary set of
cooling sequences incorporating the Saumon \& Jacobson (1999) synthetic colors for cool
WDs with pure H atmospheres provides luminosity and discovery functions in various passbands 
for a dark halo WD population consistent with the microlensing experiments (Chabrier 1999).

Current cooling models (Wood 1995; Segretain et al. 1994; Salaris et al. 1997;
Montgomery et al. 1999) appear to be reasonably consistent with each
other for WDs with $\te\simgr 4000$ K, i.e. younger than $\sim 10$ Gyr for a 0.6 $\msol$ WD.
For these temperatures, hydrogen molecular recombination is absent or negligible and the spectral energy distribution is moderatly affected by H$_2$-H$_2$ or H$_2$-He CIA (see, e.g., BSW).
The aim of the present paper is to extend cooling calculations to older WDs that are
characteristic
of the disk, spheroid or dark halo population. To
this end, we use upgraded interior physics, synthetic spectra, 
and interior-envelope relations.
A description of the various physical inputs entering the calculations, and
a comparison with previous calculations are given in
\S 2. Cooling sequences, color-magnitude and color-color diagrams in various optical and infrared
passbands are presented in \S 3, and the remaining uncertainties in the models are examined. Conclusions
are presented in Section 4.

\section{Model calculations}

In the present study, we will concentrate on the so-called DA WDs, i.e. those having either pure hydrogen atmospheres or
atmospheres with a small admixture of helium ($[N({\rm He})]/[N({\rm H})]\ll 1$).
The main reason is the fact that detectable halo WDs will have very likely hydrogen-rich atmospheres. It is indeed well
known that white dwarfs with pure helium outer layers must evolve more
rapidly than their DA counterparts at the faint end of the cooling
sequence because of the extreme transparency of these layers. Explicit
calculations, assuming such pure helium envelopes, show that
helium-atmosphere halo WDs, if they exist, would escape detection since
they reach $\mbol\simgr 19$ after $\sim 8$ Gyr (Chabrier
1997; Hansen 1998). We note, however, that even very small traces of hydrogen or
metals in their envelopes --possibly due to accretion, microscopic
diffusion, or convective dredge-up-- could change this picture
dramatically. Indeed, such traces increase significantly the opacity
of the otherwise transparent neutral helium at low effective
temperatures, slowing down the cooling of the star. If hydrogen
itself is present, even as a trace, the emergent spectrum is brought
closer to that of a DA star.

Note also that reliable calculations of pure helium model atmospheres at very low effective temperatures remain to be done. 
For instance
accurate treatments of helium pressure-ionization, as raised initially by B\"ohm et al. (1977; see also BSW and Hansen, 1998) and of
$He^-$ free-free absorption cross-section at high densities are still lacking.
 While realistic model atmospheres of non-DA WDs are
available for $\te>4000$ K (BSW, BWB) and reasonable estimates of their 
cooling timescales have been published (Wood 1995; Segretain et
al. 1994; Salaris et al. 1997), the same cannot be said at the cooler
end of the sequence. Fortunately, as mentioned above, the most probable cases of
detection of halo WDs are identified with DA stars.

\subsection{Core Physics}

\subsubsection{Equation of state}

In the present paper, we consider only WDs with carbon-oxygen cores, which restricts 
the mass range to $0.5\,\msol \le M \le 1.2\,\msol$.
We define the core as the region where the C/O plasma is fully ionized and the
electron gas is fully degenerate. The large electron conductivity ensures that the core is
very nearly isothermal.
This essentially isothermal domain encompasses typically 99.99\% of the mass of cool WDs (see, e.g., Figure 14 of Tassoul, Fontaine, \& Winget 1990). In this
region, we adopt the equation of state (EOS) described in \S2 of Segretain et al. (1994)
for the liquid and the solid phases:

\begin{equation}
U_L({x_i},\rho,T)=U_i^{id}+U_i^{ex}+U_e^{id}+U_e^x+U_{ie}
\end{equation}

\begin{equation}
U_S({x_i},\rho,T)=U_i^{th}+U_i^{anh}+U_i^{qm}+U_i^{Mad}+U_e^{id}+U_e^x+U_{ie}
\end{equation}

\noindent where the different contributions are described in Segretain et al. (1994) ($x_i=N_i/N$ denotes the number-fraction of each element, $C$ and $O$).

In the present calculations, we introduce a major improvement with respect to Segretain et al. (1994)
which is essential for cool, crystallized WDs. 
In Segretain et al. (1994) and in further calculations based on this EOS, the ion-electron screening contribution, $U_{ie}$, was taken from 
Yakovlev \& Shalybkov (1989, hereafter YS). As stressed by these authors, their calculations are valid 
only in the fluid phase, where the plasma ion coupling parameter $\Gamma
=2.275\times 10^5 \,\langle Z^{5/3} \rangle {(\rho Y_e)^{1/3}\over T} \simle 200$, where $Y_e=\langle
Z \rangle / \langle
A \rangle$ is the average electron molar fraction and $\langle Z^{5/3} \rangle
=\sum_i x_iZ_i^{5/3}$.
The arrow in Figure 1 indicates the luminosity at which the ion coupling parameter at the center of the star reaches the afore-mentioned value 200 for a 0.6 $\msol$ WD.
As shown in the figure, extrapolation of the YS fit in the solid
phase ($\Gamma > 200$) yields an increasingly inaccurate cooling sequence.
Note also that the YS fit is derived from calculations of the thermodynamic properties of a classical ionic plasma, whereas the C/O solid core is in a quantum state, with an ion diffraction 
parameter $\eta=\hbar\Omega_P/kT > 1$, where $\Omega_P$ denotes the ion 
plasma frequency (see Segretain et al. 1994).
The Segretain et al. (1994) study was primarily devoted to disk WDs, i.e., objects brighter than 
$\log L/L_\odot \sim -4.5$. This corresponds approximately to the end of the crystallization process 
in WD interiors. Halo WDs, however, are
fainter than $10^{-4.5} \lsol$ and $\Gamma > 200$ throughout a large fraction of the star.

In fully crystallized WDs, there is no further contribution to the luminosity from crystallization-induced chemical 
fractionation of C and O (see Segretain et al. 1994; Chabrier 1997; Isern et al. 1997). The only contributions to the luminosity come from the thermal reservoir of the star and from the residual gravitational contraction of the outermost layers (see e.g. Koester \& Chanmugam, 1990; D'Antona \& Mazzitelli, 1990):

\begin{equation}
dL/dt=-\int_0^M C_V {dT\over dt} dm  -\int_0^M (T{dP\over dT})_V {dv\over dt} dm.
\end{equation}

Note that the thermal energy (first term on the r.h.s. of equation (3)) is comparable to the change in gravitational energy $\Delta \Omega$, from the virial theorem.
For a substantially (entirely) crystallized WD, most of (all) the thermal energy stems from the specific heat of the quantum solid. Although for very cool WDs the thermal contribution of the central regions, where $\hbar \Omega_P/kT >> 1$, becomes increasingly small, the contribution of the outer layers, where $\hbar \Omega_P/kT \sim 1$, remains substantial.
Indeed, in cool WDs, the main contributions to the internal energy, namely the 
zero-temperature electron gas kinetic energy $U_{e}^{id}$ and the ionic electrostatic
energy $U_i^{Mad}$ in equation (2) (see Figure 1 of Segretain et al., 1994), do not depend on temperature, so that the heat capacity, and 
thus the cooling, is entirely determined by the small temperature-dependent {\it corrections} to the energy.
In fact the ion-electron screening contribution, which stems from the polarization of the electrons by the ionic field, becomes the {\it dominant} contribution to the specific heat 
at low temperature since it decreases
as $C_{V_{ie}}\propto \eta^{-1}\propto T$,
whereas the ionic crystal (Debye) contribution decreases as $C_{V_{ii}}\propto \eta^{-3}\propto
T^3$ (Potekhin \& Chabrier 2000, hereafter PC). 

For fully crystallized WDs, the central density is of the order of $\rho_c\simeq 10^6\,\gcc$, so 
that the Fermi parameter $x=p_F/m_ec=1.01\times 10^{-2}(\rho Y_e)^{1/3}>1$. Therefore we must consider the energetic contribution due to the ion-electron interaction of a polarizable, 
relativistic electron gas immersed in a quantum Coulomb crystal.
The Thomas-Fermi approximation (Salpeter 1961) is valid only in the asymptotic limit of an infinite 
ionic charge $Z\rightarrow \infty$ and is not valid for
a C$^{6+}$/O$^{8+}$ plasma (see, e.g., YS). The ion-ion and ion-electron contributions to the EOS of a 
quantum electron-ion solid plasma for a finite ionic charge under the conditions of interest have been 
calculated recently by Potekhin \& Chabrier (2000). To the best of our knowledge these 
are the only available calculations for such plasma conditions. Figure \ref{fig1-cbfs-bw.eps}
compares the evolution of a 0.6 $\msol$ crystallized WD for the same $\te$-$T_c$ condition 
(as described below) with the solid EOS and specific heat incorporating (i) the proper ion-electron 
screening treatment, and (ii) an extrapolation of the YS fit. As shown, the extrapolation yields 
increasingly shorter cooling times for WDs older than $\sim 8$ Gyr. 

The contribution of crystallization along evolution is treated as described
in \S 4 of Segretain et al. (1994), from the evolution of the binding energy, $dB(T)/dt$. Recent complete
evolutionary calculations by Montgomery et al. (1999) confirm the validity of this method for cool WDs, where there is no neutrino or nuclear luminosity.
The crystallization of $^{22}$Ne, which yields an azeotropic diagram (Segretain
\& Chabrier
1993), and was originally thought to produce a significant time-delay (Isern et
al 1991;
Segretain et al. 1994) has not been considered in the present calculations. Indeed,
consistent calculations of the three-body Ne/C/O phase diagram (Segretain 1996) show
that
when Ne-crystallization sets in, a substantial fraction of oxygen has already crystallized,
yielding an O-rich core, so that the remaining amount of energy due to $^{22}$Ne
crystallization becomes fairly small. At the crystallization luminosity found in Segretain et al. (1994), the induced time delay is at most $\sim 0.3$ Gyr, well within the
remaining uncertainties in our calculations (phase diagram, opacities, etc).
In any event, since we are presently mainly interested in old WDs originating from
low-metallicity progenitors, the Ne abundance is likely to be too small
to have any measurable effect.

\subsubsection{Initial carbon-oxygen profiles}

Substantial uncertainty remains on the $^{12}C(\alpha,\gamma)^{16}O$ reaction which determines the 
final state of the He-burning AGB phase and thus the initial C/O abundance-profile of the WD. Modern 
experimental data and updated values of the astrophysical $S$-factor (see, for example, Arnould et al. 1999) 
suggest a rate about a factor 1.5 to 2 larger than that of Caughlan \& Fowler (1988,CF88), 
closer to the previous Caughlan et al. (1985) value. This yields a substantially O-enriched initial 
profile in the WD core (see e.g. Salaris et al. 1997). We have adopted these initial profiles in our 
calculations. However, in order to examine the uncertainties due to the interior composition, we have 
also conducted
calculations for a 0.6 $\msol$ WD with an initial C/O distribution resulting from the CF88 lower rate (see Figure 2 of Salaris et al. 1997).

\subsection{The Luminosity-Central Temperature Relation}

The binding energy method that we use to compute cooling ages relies on
the availability of $L$-$T_c$ relations which govern the cooling rates
of WDs. These relations are particularly sensitive to the constitutive
physics of the outer partially ionized, partially degenerate envelope
which connects the nearly isothermal core of a WD to its surface (the
atmospheric layers). Previous calculations based on this method
(Segretain et al. 1994 and references therein; Salaris et al. 1997) used $L$-$T_c$ relations provided by independent model calculations such as, for
example, those of Wood \& Winget (1989) or Wood (1995). This approach necessarily
introduces some inconsistencies in the calculations: the chemical
composition of the core of the models used to derive the $L$-$T_c$
relation differs from the variable core composition of the nearly
isothermal structure that undergoes phase separation, and the constitutive
physics is generally different in the two sets of models. Moreover, for
masses not directly available from independent models, $L$-$T_c$ relations
were {\it scaled} on the mass, which, at best, provides a rough estimate of
the correct relations.

We have improved on this front in the present paper by computing $L$-$T_c$ relations based on state-of-the-art constitutive physics and by considering several individual masses. We are still left with the inherent
inconsistency of the binding energy approach due to the fixed interior composition in the calculation of the $L$-$T_c$ relation, but this shortcoming is largely offset by our ability to
describe in accurate details the physics of crystallization (e.g., the
phase diagram) or the radiative and conductive opacities at high density,
or by the afore-mentioned uncertainties in the stratified C/O profile. Note
also that the treatment of crystallization-induced fragmentation can be easily implemented in the binding energy method, whereas it is a complicated task to include it into a standard evolutionary code. At this level, the binding energy method shows an appreciable advantage.
Moreover, as mentioned above, the validity of
this method for cool WDs has been assessed recently by comparisons with complete evolutionary calculations (Montgomery et al. 1999).

The $L$-$T_c$ relations for various masses in the range of interest were computed with an upgraded version of
the stellar model building code briefly described in Brassard \&
Fontaine (1994, 1997). These are {\it full} stellar models that describe
the complete structure of a static WD from the center to the high
atmosphere ($\tau_R \sim 10^{-6}$, where $\tau_R$ denotes the Rosseland optical depth). The envelope calculation incorporates the SCVH EOS for H and He and the
Fontaine, Graboske, \& Van Horn (1977) EOS for carbon and oxygen. The
radiative opacities include the OPAL 1996 data (Iglesias \& Rogers
1996) complemented at low temperatures by the Rosseland
opacities of H and He computed with the model atmosphere code of BSW down to 1500 K for H and 2500 K for He. These latter opacities include the complete CIA processes (see BSW). For the conductive opacities, we use a large table incorporating the Hubbard \& Lampe (1969) and
Itoh et al. (1983, 1993) calculations (Brassard \& Fontaine 1994), which covers the entire density and 
temperature range relevant to the present calculations. Convection is described with the
standard mixing-length theory. As shown in Tassoul et al.
(1990), the $L$-T$_c$ relationship is insensitive to the assumed convective treatment, since 
convection is essentially adiabatic when it breaks through the degenerate core.

We note that Hansen (1998, 1999) has recently emphasized the importance
of treating in detail the atmospheric layers in the context of the
evolution of very cool WDs. As discussed initially by Fontaine
\& Van Horn (1976; see also Tassoul et al. 1990), the most important
consequence of such a detailed description of the atmosphere on the
cooling is its effect on the location of the base of the
convection zone in the deeper, optically-thick envelope. Indeed, below a luminosity 
$\log (L/\lsol) \sim -3.7$ ($\te\simeq 5500-6500$ K), the base of the
hydrogen superficial convection zone reaches into the degenerate core,
thus coupling, for the first time during the evolution, the atmospheric
layers with the central thermal reservoir. Since, by then, the
stratification of the envelope is fully convective and highly adiabatic,
small changes at the base of the atmosphere produce corresponding 
changes at the base
of the convection zone. A correct determination of the base of that zone
is thus essential to calculate an accurate $L$-$T_c$ relation for
cool WDs.

It is important to stress that, although the atmospheric structures entering the present stellar
models are gray, they take into account
the feedback effect of convection on the atmospheric structure and thus are not based on a Rosseland mean all the way through, unlike those used in all previous
evolutionary calculations, with the exception of Hansen's (1998; 1999) computations. These
modified model atmospheres, to be described elsewhere (Brassard \& Fontaine, in preparation), reproduce almost exactly the stratification of a detailed model atmosphere (from, e.g., BSW in the present context) at large optical depths. They reproduce well, in particular,
the main effect due to nongrayness, namely the upward shift of the
convection zone. This phenomenon is well known in stellar atmosphere
theory and has been described, in a white dwarf context, by B\"ohm et
al (1977) among others.
Therefore, the boundary conditions provided by the
detailed model atmosphere at $\tau_R = 100$ are
essentially the same as those provided by the gray model with convective
feedback. This is what matters for the cooling; for nearly
identical boundary conditions at the base of the convective atmosphere,
we find the same location of the base of the full envelope convection
zone, and hence the same value of $T_c$. In contrast, standard gray atmosphere
stratification (not taking into account the feedback due to convection)
leads to an incorrect determination of the base of the convective zone, overestimating its penetration and thus yielding a faster cooling (see \S 3.1 below).

A central parameter in the envelope calculation, and in the resulting cooling time, is
the amount of hydrogen and helium present in the envelope. For the present calculations, we have used DA white dwarf standard ``thick'' layers, $\log q(H)=-4.0$, 
$\log q(He)=-2.0$, where $q(X)=M(X)/M_\star$ denotes the mass fraction of 
element $X$. In practice, the masses of the hydrogen and helium layers on top of the degenerate C/O interior depend on the WD mass, through the AGB and post-AGB evolutionary phase (see e.g. Bl\"ocker et al., 1997). Note, however, that (i) calculations during the AGB phase strongly depend on ill-constrained parameters (overshooting, mass loss,...) and (ii) present calculations consider the evolution of solar-metallicity AGB stars, whereas the progenitors of very cool WDs are metal-depleted. Given these uncertainties in the exact amount of $q(H)$ and $q(He)$, we elected to conduct our calculations with the afore-mentioned "standard" values for 0.6 $\msol$ WD (see Fontaine \& Wesemael 1997 for a general discussion of this unsettled issue).
This uncertainty in the exact amount of H and He in WD envelopes certainly represents one of the major uncertainties in present WD cooling calculations. Some scaling relations, however, can be used. Indeed,
the effect of the thickness of the helium layer on the cooling time 
has been examined in detail by Tassoul et al. (1990), Wood (1992) and Montgomery et al. (1999, \S 5.1). Thicker helium layers  result in
more transparent envelopes (since $\kappa_{\rm He}\ll \kappa_{\rm Carbon}$ under WD conditions), 
decreasing the temperature gradient between the core and the surface, so that the central 
temperature decreases faster with decreasing luminosity.
Therefore, models with thicker He layers are younger for a given mass and luminosity, with a $\sim 0.75$ Gyr decrease in the age for each order of magnitude in $M_{He}$ (Montgomery et al. 1999).

\subsection{Model atmospheres and photometric colors}

The observable properties of the cooling WD, such as the emergent spectrum and the 
photometric colors are obtained with atmosphere models.
The present calculations include the colors and bolometric corrections for pure hydrogen atmospheres
calculated by Bergeron et al. (1995a) above 4000 K, extended down to $\te=1500$ K by Saumon \& Jacobson (1999).
As mentioned previously, for very cool WDs ($\te \wig < 5000$ K), molecular hydrogen becomes stable
and the main source of opacity in the infrared is the CIA by H$_2$. This opacity forces the stellar 
flux to emerge at shorter wavelengths, with a peak near 1$\,\mu$m.
This increased flux in the $R$ and $I$ bands and decreased flux
in infrared results in increasingly blue color indices for cooler WDs (Hansen 1998, 1999; SJ).

Although similar to those of Hansen (1998, 1999), the present calculations include
a more detailed treatment of the microphysics entering the atmosphere (BSW and SJ). Indeed, an
important feature in these cool and dense atmospheres
is the effect of the surrounding particles on the partition function
of an atom, which eventually leads to the pressure-ionization of hydrogen. 
These modified internal partition functions imply a different
(non-ideal) ionization equilibrium, in particular for the abundances of H$_2$,
H$_2^+$ and H$_3^+$. This in turn modifies the abundance of free electrons, and thus
of H$^-$ ions, a dominant source of opacity in the optical.
The atmosphere models of SJ were used in the range $1500\le \te \le 4000$ K 
and $7.5 \le \log g \le 9.0$,
complemented by BSW for $4000\le \te \le 10000$ K, with a pure hydrogen composition.
These synthetic spectra and atmosphere models successfully reproduce
spectroscopic and photometric observations of cool H-rich (DAs) WDs above 4000 K (Bergeron et al. 1997).

Optical colors for cool WDs with a small but non-zero $N({\rm He})/N({\rm H})$ 
composition  ratio will be only slightly different
($\simle 0.2$ mag) from the colors of pure hydrogen atmospheres presented here
(see BSW; Bergeron et al. 1997).
The color indices of Table 2 of BSW can then be used above 4000 K for mixed composition atmospheres.

\section{Results}

\subsection{Cooling curves}

For the purposes of comparisons, we define a set of reference models which include the physics
described in the previous section, including crystallization-induced fragmentation (see below), 
with the Salaris et al. (1997) high-rate stratified initial C/O profiles.
The $L$-$T_c$ relations that we used for these reference models
have been computed as described above, from full static stellar models
with chemical layering defined by $\log q({\rm H})=-4.0$ and 
$\log q({\rm He})=-2.0$. The photometric colors are taken from pure-H
atmosphere calculations.

\subsubsection{Comparison with existing calculations}

We first compare, in Figure \ref{fig2-cbfs-color.ps}, our $L$-$T_c$ relation for a 0.6 $\msol$ 
WD with the one obtained by Wood (1995), Montgomery et al. (1999, Table 1) and another
one kindly provided by Hansen (1999) for the same values of 
$q({\rm H})$ and $q({\rm He})$. The four curves agree reasonably well,
but the relatively small differences shown here translate into
significant differences on the cooling time (see below). Unfortunately,
it is currently impossible to account in detail for the differences in
the $L$-$T_c$ relations illustrated in Figure 2;
differences in the constitutive physics used by the various groups are
probably at the origin of most of the deviations. The $L$-$T_c$ relation
is particularly sensitive to the opacity profile throughout the star, so
slightly different implementations of the conductive and radiative
opacities, as well as the use of different generations of these data,
could very well account for most of the discrepancies. As expected, the Wood (1995) and the Montgomery et al. (1999) relations are very similar, since they rely on the same input physics, yielding similar cooling sequences.

As mentioned in \S 2.2., the change of slope observed in all three curves is a well known
phenomenon and is due to
convection breaking into the degenerate thermal reservoir at
sufficiently low luminosities. When that occurs, there is a flattening
of the temperature gradient between the core and the surface due to the
larger efficiency of convection as compared to radiation. Initially,
this leads to a slowdown of the cooling process as an excess energy is
liberated through the more transparent convective envelope, but this is
rapidly followed by a phase whereby convection speeds up the cooling
process as compared to the case of purely radiative models (see Figure 3
of Tassoul et al. 1990 and related discussion). For luminosities lower than that of the breaks in slope shown
in Figure 2, the $L$-$T_c$ relation becomes sensitive, among other things,
to the details of the atmosphere. From this point of view, the Hansen
(1999) calculations and our own improve upon those of Wood (1995) or Montgomery et al. (1999)
because the latter ones rely on a standard gray atmosphere strategy which
overestimates the penetration of the superficial convection zone and
leads to a value of $T_c$ slightly lower than it should be, as discussed previously.

On the other hand, for luminosities larger than those of the changes of
slope, it is well established that the
relation is completely insensitive to the stratification of the upper
envelope and, in particular, to the details of the atmosphere (see
Tassoul et al. 1990 for a complete discussion). The
discrepancies shown in Figure 2 between the three curves for these higher
luminosities are then the explicit proof that there are indeed
significant differences in the implementation/calculation of the
constitutive physics between the three groups, notably at the level of
the conductive opacities. So, to a certain extent, the differences found
in Figure 2 result from the fact that we are comparing models with
different physical inputs.

Figure \ref{fig3-cbfs-color.ps} illustrates the consequences of these differences on the cooling 
time of the star. For each $L$-$T_c$ relation, we show two cases:
(i) taking into account the chemical fractionation of C and O at 
crystallization (rightmost curves), (ii) ignoring this process 
(leftmost curves). Interestingly enough, we find that in the phases of interest for the present study $[\llsol < -4.5]$), our results (solid curves) agree reasonably well with
those obtained when using the $L$-$T_c$ relation provided by Hansen (long-dashed curves). The
results derived on the basis of the Wood (1995) (or Montgomery et al. 1999) $L$-$T_c$ relation
(dotted curves) provide, in comparison, shorter cooling
timescales, as anticipated from the discussion above. The differences illustrated in Figure \ref{fig3-cbfs-color.ps} simply amplify the
differences already observed in the $L$-$T_c$ relations of
Figure \ref{fig2-cbfs-color.ps}.

Another interesting comparison is provided by the short-dashed curve in
Figure \ref{fig3-cbfs-color.ps} which presents the cooling curve obtained by Hansen
himself (1999), which is supposed to include the complete treatment of crystallization and thus must be compared with the right long-dash curve.
The difference between the two curves is noticeable ($\simgr 0.5$ Gyr) and can not be totally ignored. Since such a comparison eliminates the effect of the energy
transfer problem (the $L$-$T_c$ relations are the same), the remaining
discrepancies must be blamed mostly on the different treatments of the
thermodynamics of the ionized interior, outlined in \S2.1.1. Since Hansen has not detailed his interior EOS, it is not possible to pursue this point further.
Note, however, that the treatment of crystallization in Hansen's calculations has been demonstrated to be incorrect (Isern et al. 2000). We thus disagree with him about the
importance of chemical fractionation at crystallization, as discussed in the
next subsection. Note also that apparently Hansen's calculations do not extend beyond $\log L/\lsol\sim -5.0$, which 
corresponds to an age of $t\sim 12$ Gyr for a 0.6$\,M_\odot$ WD.

\subsubsection{Effects of internal composition and crystallization}

Figure \ref{fig3-cbfs-color.ps} illustrates also the effect of initial C/O stratification on the 
cooling time for a 0.6 $\msol$ H-atmosphere WD, as examined in detail by Salaris et al. (1997). The rightmost
dot-dash curve displays the cooling sequence of our reference model with an 
initial C/O profile resulting from a low (CF88) $^{12}C(\alpha,\gamma)^{16}O$ reaction rate, to be compared with the sequence obtained with a high rate (Caughlan et al. 1985) induced profile (rightmost solid line). 
A larger rate yields a larger initial oxygen-enriched core and affects the cooling in
several ways (Segretain et al. 1994; Salaris et al. 1997): (i) the larger oxygen content corresponds to a smaller heat capacity $[C_V\propto \sum_i (X_i/A_i)]$ and thus to a 
faster cooling prior to crystallization, (ii) the gravitational energy release at crystallization,
$\Delta E\propto {\Delta \rho \over \rho}Mg$, is smaller because of the spindle form of the 
phase diagram (Segretain \& Chabrier 1993; Segretain et al. 1994; Chabrier 1997) but 
(iii) the total amount of oxygen to be differentiated is larger [cf. (i)],
and (iv) 
crystallization occurs earlier in the evolution since oxygen crystallizes at a higher 
temperature than carbon. Whether this corresponds to a larger {\it effective} temperature 
depends on the $L$-$T_c$
(and thus $\te$-$T_c$) relation for each WD mass (Figure \ref{fig2-cbfs-color.ps}). As shown by 
Salaris et al. (1997), effects (ii) through (iv) more or less compensate for both reaction rate induced profiles, and
the time delay induced by crystallization $\Delta \tau=\Delta E/L$ is about the same for these 
two stratified profiles. The difference in cooling times thus stems primarily from the 
available heat content [point (i)]. For example, at a luminosity $\log (L/\lsol)=-4.5$ (resp. -5.0),
our reference 0.6 $\msol$ model has an age of $t=10.3$  (resp. 12.7) Gyr
whereas the same luminosity corresponds to an age $t=10.7$ (resp. 13.3) Gyr for a low-rate 
initial profile (see Figure \ref{fig3-cbfs-color.ps}), confirming the Salaris et al. (1997) analysis.
This illustrates the present uncertainty in cooling times due to uncertainties in 
the $^{12}$C$(\alpha,\gamma)^{16}$O reaction rate and induced WD initial internal composition.

Another uncertainty affecting the internal composition is the exact shape of the C/O 
crystallization diagram. In the present calculations, we use the Segretain \& Chabrier (1993) 
spindle diagram, calculated within the framework of the density-functional theory of freezing. 
The Barrat, Hansen, \& Mochkovitch (1988) calculations were based on previous (obsolete) 
values of the plasma parameter $\Gamma$ at crystallization but would yield similar results if 
updated (see Segretain \& Chabrier 1993),
whereas the Ichimaru, Iyetomi, \& Ogata (1988) azeotropic C/O diagram is demonstrably erroneous 
(DeWitt, Slattery, \& Chabrier 1996). Figures \ref{fig3-cbfs-color.ps} and and  \ref{fig4-cbfs-color.ps} illustrate the effect of chemical 
fractionation at crystallization, due to the difference of abundance of carbon and oxygen in the 
fluid and in the solid phase, on a 0.6 $\msol$ WD. This induces a variation of chemical potential at constant volume 
and temperature (Chabrier 1997) which provides an additional source of energy (Mochkovitch 1983; 
Garc\'\i a-Berro et al. 1988; Segretain et al. 1994). Although this energy amounts to only 
$\sim 1\%$ of the binding energy of the star, it is released at a low luminosity and thus 
lengthens appreciably the lifetime of the star
(Chabrier 1997; Isern et al. 1997).  For example, a luminosity $\log (L/\lsol)=-4.5$ (resp. -5.0)
for our stratified profile corresponds to an age $t=8.8$ (resp. 11.4) Gyr 
if fractionation is ignored and $t=10.3$ (12.7) Gyr if it is taken into account.
As seen in figure \ref{fig3-cbfs-color.ps}, once crystallization has proceeded throughout the entire 
star, the time delay remains constant, and amounts from $\sim 1$ Gyr for the least (0.5 $\msol$) and most (1.2 $\msol$) massive WDs, to a maximum value of about 1.5 Gyr for our reference 0.6 to 0.8 $\msol$ 
WDs. The effect thus remains substantial and must be properly taken into account in WD cooling theory.
This is at odds with Hansen's (1999) results, as explained in \S 3.1.1. As seen in Figure \ref{fig3-cbfs-color.ps}, crystallization in our calculations occurs at a later age and a fainter luminosity than in the Wood (1995) or Montgomery et al. (1999) calculations. This reflects the faster cooling for Wood and Montgomery's calculations, as discussed in \S 3.1.1. This results in a smaller crystallization-induced delay in this latter case since $\Delta \tau \propto 1/L$.

Although the crystallization model for
binary ionic mixtures has not yet been verified observationally in WDs,
it has been studied for a long time by geophysicists
(see, for example, Loper 1984; Buffett et al. 1992). Although the nature of the plasma
(or alloy) is different, the {\it physics} of the process (thermodynamics and energy transport) is exactly the same. Note, however, that these crystallization-induced delays represent upper limits. Indeed, the present calculations assume complete mixing of the C-enriched fluid layers, i.e. a maximum efficiency for this process. This is supported by the fact that the mixing instability timescale is much shorter than the evolutionary timescale (Mochkovitch, 1983).

Figure \ref{fig4-cbfs-color.ps} displays the cooling sequences $\mbol(t)$ for different masses for our reference model calculations.
 
\subsection{Cooling times}

Figure \ref{fig5-cbfs-bw.eps} displays the mass-$\te$ relation, or equivalently the radius- or 
surface gravity-$\te$ relation (see Table 1-5),
for several constant WD cooling times.
Before crystallization sets in, massive WDs evolve more slowly, because of their greater energy 
content ($C_V$) and their smaller radiative surface areas\footnote{Remember the objects under consideration are cool enough so that the neutrino luminosity is completely negligible.}. Since they are hotter and brighter at a given $T_c$ than less-massive WDs, and 
since crystallization occurs always at the same internal temperature $T_c$, massive WDs 
crystallize earlier and at a higher $\te$ and $L$. At this stage, the crystallized core enters the 
Debye cooling regime ($C_V\propto T^3$) and cools more quickly. This crystallization process 
causes the bending on the cooling times downward.

Figure \ref{fig6-cbfs-bw.eps} displays the same constant cooling times as a function of absolute $\mv$ magnitude. 
The bulk of all hydrogen-atmosphere WDs, those with $m \simle 0.8\,\msol$ (assuming a WD mass distribution similar to the one observed in the disk), remains brighter than 
$\mv\simeq 18$ after 14 Gyr.

\subsection{Color-magnitude and color-color diagrams}

Figures \ref{fig7-cbfs-color.ps} and \ref{fig8-cbfs-color.ps} display cooling sequences in $\mv$ vs. $(V-I$) 
and $\mj$ vs. $(J-K)$ diagrams, respectively.
WDs with a small admixture of helium in the atmosphere will have similar colors (see Tables 1 
and 2 of BSW). The triangles illustrate the coolest disk WDs  identified spectroscopically as H-rich atmosphere WDs by Leggett, Ruiz, \& Bergeron (1998). Collision-induced absorption by H$_2$ at $\te\sim 4500$-5000 K causes the 
turnover to bluer $J-K$ at $\mj \sim 13$-$15$, depending on the mass. At these temperatures, the 
effect is still modest and a large fraction of the flux emerges longward of 1 $\mu$m 
(see Figure 5 of BSW). As $\te$ decreases, however,
the CIA becomes stronger, causing a turnover in the optical colors near $\mv \sim 16$-$17$ at 
later stages of the evolution.
The 12 Gyr curve forms a loop in the $\mv$ vs. $(V-I)$ diagram. This stems from the very 
different cooling rates of WDs for different masses. As mentioned above, less massive WDs cool faster initially 
because of their smaller heat content, whereas massive WDs crystallize earlier and then 
enter the rapid Debye cooling regime. The
combination of these two effects yields the slowest cooling rate for $\sim 0.8\,\msol$ WDs 
after $\sim 8$ Gyr, whereas both the least and most massive objects cool faster, 
reaching fainter magnitudes and bluer colors at younger ages. 

Figure \ref{fig9-cbfs-color.ps} shows a color-color diagram comparing optical and near-IR colors
for a 0.6 and a 1.2 $\msol$ H-atmosphere WD.
Note that gravity does not affect the qualitative behavior of the diagram. Indeed increasing 
gravity has an effect similar to decreasing the effective temperature on the spectrum (see, e.g., Saumon 
et al. 1994). Naturally the age-dependence is different, as shown on the figure by the solid circles and squares, respectively.
As already mentioned in Chabrier (1999), we note from the previous figures the strong 
dependence of colors upon age for $t\simgr 12$ Gyr. If the gravity, i.e., the mass of such a WD, 
can be determined independently, this provides a powerful tool to determine the age of the Galactic halo.

Tables 1-5 give the characteristic properties of the present WD
H-rich atmosphere cooling sequences in various broadband filters for the mass-range characteristic 
of WDs with C/O cores.

\section{Conclusion}

We have computed evolutionary sequences for cool ($\te < 10\,000$ K) 
pure hydrogen atmosphere WDs, although the calculations can be applied 
to WDs with a small admixture of He in the atmosphere as well, using 
appropriate color indices (BSW). These models are primarily aimed at 
identifying old, cool WDs in the Galactic old disk, spheroid or dark 
halo, or in globular clusters once fainter detection limits can be achieved.
We have first improved upon the equation of state of the fully ionized
interior to calculate accurate cooling sequences for crystallized WDs, 
namely objects fainter than $\log (L/\lsol) \sim -4.5$ ($t \simgr 9$ Gyr
for a 0.6 $\msol$). These WDs are in a quantum regime. The 
temperature-dependent contributions, in particular the ion-electron 
energy of a quantum, relativistic ion-electron plasma, are the only
sources of specific heat and thus determine entirely the thermal 
reservoir to be radiated into space, even though they provide a 
negligible contribution to the internal energy. We have next obtained
improved $L$-$T_c$ relations which govern the rate of cooling of
white dwarfs. The envelope and atmosphere calculations include 
state-of-the-art radiative opacities down to $\te= 1500$ K for pure
hydrogen. We have examined and quantified the uncertainties in these 
cooling sequences arising from (i) the initial C/O stratification of 
the WD, which follows from the $^{12}$C$(\alpha,\gamma)^{16}$O reaction 
rate, (ii) the $L-T_c$ relation (the core-surface boundary condition), and 
(iii) the crystallization-induced gravitational energy release. The 
uncertainty in the $^{12}$C$(\alpha,\gamma)^{16}$O reaction rate 
translates into a $\simle 0.5$ Gyr difference in cooling time. The 
uncertainty due to different $L$-$T_c$ relations when using accurate 
boundary conditions yields substantial differences in the range 
$(L/\lsol) \sim 10^{-4.0}$ -- $10^{-4.5}$, when, for the first time, the 
superficial convection zone reaches the degenerate core. This is 
probably due to the detailed physics entering the core (EOS) and 
envelope (EOS, $\nabla_{\rm rad}$, $\nabla_{\rm cond}$) calculations. 
These differences, however, vanish almost entirely below this limit, i.e., 
for $t\simgr 10$ Gyr. The maximum time delay induced by chemical fractionation
at crystallization amounts to $\sim 1$ to 1.5 Gyr, depending on the WD 
mass, and represents one of the major uncertainties in dating very cool WDs from their observed luminosity. Future laser-driven experiments on the 
crystallization of dense plasmas should shed light on this complicated 
physics problem, which bears major consequences in the present 
astrophysical context. The other major uncertainty in WD cooling stems from the ill-determined mass fraction of hydrogen and helium in the external envelope, in particular for old WDs originating from metal-depleted progenitors.

We find noticeable differences ($\simgr 1$ Gyr) between our and 
Hansen's (1999) cooling sequences. Part of it very likely stems from the different treatments of the interior EOS and envelope EOS and opacities, but also from the demonstrably underestimated crystallization-induced time delay in Hansen's calculations. A $\sim 0.5$-1 
mag difference in colors
occurs also for $V-I\simle 1$ between Hansen's and our calculations, which most likely  arises
from details of the calculation of the atmosphere models, including the non-ideal effects 
(see Figure 3 of SJ).
This shows that, although a consistent general theory for the cooling of cool WDs is emerging, 
work remains to be done to reach more robust results. The theory is still very uncertain for 
very cool helium-rich atmosphere WDs.
A correct calculation of helium pressure ionization and of the He$^-$ free-free absorption 
cross-section at high densities in the atmosphere of these objects remains to be done. 
Attention must also be devoted to dense atmospheres including a trace of metals since the 
subsequent increase in opacity will affect dramatically the cooling of the star. Work in this 
direction is in progress.

Meanwhile, the present calculations should provide what we believe to be presently the most 
accurate calculations of very cool white dwarf sequences.
They provide a useful basis to search for and to identify faint, old WDs either in the field or 
in globular clusters. By allowing the determination of the mass and age of possible halo WDs 
(Hodgkin et al. 2000, Ibata et al. 2000), they will also provide important constraints on the age of the 
Galactic disk and halo, and on the Galaxy initial mass function.

\bigskip

Note: The present models in various filters are available upon request to Gilles Chabrier (chabrier@ens-lyon.fr).
\begin{acknowledgements} We thank
M. A. Wood and B. Hansen for kindly providing models and results from their calculations, M. Hernanz for providing tables of the stratified profiles
and A. Potekhin for stimulating discussions.
This work was supported in part by NSF grant AST-9731438 to D.S.
\end{acknowledgements}

\vfill
\eject

{}

\vfill
\eject

\begin{table*}
\caption{Cooling sequence for a 0.5 $\msol$ white dwarf of our reference model.
The age is in Gyr, $T_{eff}$ is in K, the surface gravity $g=GM/R^2$ in cgs.
$M_{bol}=-2.5\, \log (L/\lsol)+4.75$.
 The VRI magnitudes are in the
Johnson-Cousins system (Bessell, 1990), JHK in the CIT system (Leggett, 1992). Absolute calibration from the observed spectrum
of Vega (Mountain et al. 1985).
}
\bigskip
\begin{tabular}{lccccccccccc}
\hline\noalign{\smallskip}

 age & $T_{eff}$ &  $\log \,g$ &  $M_{bol}$ &$M_B$ &$M_V$ &$M_R$ &$M_I$ &$M_J$ &$M_H$ & $M_K$ \\
\noalign{\smallskip}
\hline\noalign{\smallskip}
 1& 7333 & 7.837 & 12.96 & 13.53 & 13.17 & 12.94 & 12.67 & 12.48 & 12.33 & 12.33 \\
 3& 5263 & 7.863 & 14.45 & 15.41 & 14.70 & 14.25 & 13.79 & 13.28 & 13.01 & 12.92 \\
 6& 4509 & 7.872 & 15.15 & 16.37 & 15.48 & 14.92 & 14.36 & 13.78 & 13.64 & 13.63 \\
 8& 3911 & 7.874 & 15.78 & 17.16 & 16.12 & 15.47 & 14.82 & 14.31 & 14.33 & 14.40 \\
 10& 3243& 7.875 & 16.59 & 17.89 & 16.72 & 16.00 & 15.40 & 15.46 & 15.67 & 15.90 \\
 12& 2578& 7.877 & 17.59 & 18.65 & 17.33 & 16.68 & 16.54 & 17.10 & 17.27 & 17.79 \\
 13& 2233& 7.877 & 18.22 & 19.12 & 17.70 & 17.15 & 17.51 & 18.18 & 18.23 & 19.10 \\
 14& 1938& 7.877 & 18.83 & 19.61 & 18.10 & 17.66 & 18.63 & 19.34 & 19.23 & 20.58 \\
 15& 1670& 7.877 & 19.48 & 20.16 & 18.55 & 18.25 & 19.94 & 20.67 & 20.34 & 22.31 \\
\hline
\end{tabular}
\end{table*}

\begin{table*}
\caption{Same as Table 1 for 0.6 $\msol$.
}
\begin{tabular}{lccccccccccc}
\hline\noalign{\smallskip}
 age & $T_{eff}$ &  $\log \,g$ &  $M_{bol}$ &$M_B$ &$M_V$ &$M_R$ &$M_I$ &$M_J$ & $M_H$ & $M_K$ \\
\noalign{\smallskip}
\hline\noalign{\smallskip}
 1& 8273 & 8.001 & 12.67 & 13.24 & 12.95 & 12.76 & 12.56 & 12.49 & 12.38 & 12.41 \\
 3& 5726 & 8.023 & 14.31 & 15.11 & 14.50 & 14.11 & 13.71 & 13.28 & 13.03 & 12.96 \\
 6& 5091 & 8.032 & 14.84 & 15.85 & 15.10 & 14.62 & 14.14 & 13.60 & 13.36 & 13.30 \\
 8& 4709 & 8.035 & 15.19 & 16.34 & 15.50 & 14.97 & 14.44 & 13.87 & 13.70 & 13.68 \\
 10& 4024& 8.038 & 15.88 & 17.22 & 16.22 & 15.58 & 14.95 & 14.40 & 14.40 & 14.46 \\
 12& 3190& 8.039 & 16.88 & 18.14 & 16.96 & 16.25 & 15.68 & 15.84 & 16.06 & 16.32 \\
 13& 2813& 8.039 & 17.44 & 18.57 & 17.31 & 16.62 & 16.27 & 16.73 & 16.93 & 17.34 \\
  14& 2355& 8.040 & 18.21 & 19.14 & 17.76 & 17.18 & 17.38 & 18.02 & 18.11 & 18.85 \\
 15& 1938& 8.040 & 19.05 & 19.80 & 18.30 & 17.88 & 18.87 & 19.58 & 19.47 & 20.81 \\
\hline
\end{tabular}
\end{table*}

\begin{table*}
\caption{Same as Table I for 0.8 $\msol.$
}
\begin{tabular}{lccccccccccc}
\hline\noalign{\smallskip}
 age & $T_{eff}$ &  $\log \,g$ &  $M_{bol}$ &$M_B$ &$M_V$ &$M_R$ &$M_I$ &$M_J$ & $M_H$ & $M_K$ \\
\noalign{\smallskip}
\hline\noalign{\smallskip}
 1& 10394& 8.316 & 12.14 & 12.79 & 12.57 & 12.49 & 12.44 & 12.53 & 12.51& 12.58 \\
 3& 7253 & 8.325 & 13.73 & 14.30 & 13.94 & 13.69 & 13.41 & 13.22 & 13.06 & 13.06 \\
 6& 5806 & 8.331 & 14.71 & 15.49 & 14.89 & 14.50 & 14.12 & 13.71 & 13.48 & 13.41 \\
 8& 5464 & 8.335 & 14.98 & 15.85 & 15.19 & 14.77 & 14.35 & 13.88 & 13.64 & 13.57 \\
 10& 5053& 8.337 & 15.33 & 16.33 & 15.58 & 15.10 & 14.62 & 14.09 & 13.88 & 13.84 \\
12& 4162& 8.340 & 16.18 & 17.42 & 16.47 & 15.86 & 15.26 & 14.76 & 14.77 & 14.83 \\
 13& 3538& 8.340 & 16.88 & 18.16 & 17.07 & 16.39 & 15.76 & 16.65 & 15.83 & 16.00 \\
14& 2768& 8.341 & 17.95 & 19.02 & 17.77 & 17.10 & 16.82 & 17.31 & 17.51 & 17.93 \\
 15& 1713& 8.341& 20.02 & 20.68 & 19.11 & 18.80 & 20.37 & 21.10 & 20.80 & 22.64 \\
\hline
\end{tabular}
\end{table*}

\begin{table*}
\caption{Same as Table 1 for 1.0 $\msol.$
}
\begin{tabular}{lccccccccccc}
\hline\noalign{\smallskip}
 age & $T_{eff}$ &  $\log \,g$ &  $M_{bol}$ &$M_B$ &$M_V$ &$M_R$ &$M_I$ &$M_J$ & $M_H$ & $M_K$ \\
\noalign{\smallskip}
\hline\noalign{\smallskip}
1& 13680& 8.621 & 11.47 & 12.55 & 12.43 & 12.48 & 12.53 & 12.75 & 12.77& 12.84 \\
3& 9609 & 8.629 & 13.02 & 13.66 & 13.41 & 13.27 & 13.16 & 13.21 & 13.16 & 13.22 \\
 6& 6794 & 8.633 & 14.54 & 15.15 & 14.72 & 14.43 & 14.13 & 13.89 & 13.70 & 13.68 \\
 8& 5761 & 8.636 & 15.26 & 16.04 & 15.43 & 15.04 & 14.66 & 14.24 & 14.02 & 13.95 \\
 10& 5389& 8.639 & 15.56 & 16.45 & 15.77 & 15.33 & 14.91 & 14.44 & 14.21 & 14.14 \\
 12& 4045& 8.640 & 16.81 & 18.03 & 17.06 & 16.45 & 15.85 & 15.45 & 15.51 & 15.61 \\
 13& 2107& 8.641 & 19.64 & 20.43 & 19.01 & 18.53 & 19.18 & 19.85 & 19.84 & 20.88 \\

\hline
\end{tabular}
\end{table*}

\begin{table*}
\caption{Same as Table 1 for 1.2 $\msol.$
}
\begin{tabular}{lccccccccccc}
\hline\noalign{\smallskip}
 age & $T_{eff}$ &  $\log \,g$ &  $M_{bol}$ &$M_B$ &$M_V$ &$M_R$ &$M_I$ &$M_J$ & $M_H$ & $M_K$ \\
\noalign{\smallskip}
\hline\noalign{\smallskip}
1& 21950& 8.995 & 10.15 & 12.27 & 12.34 & 12.45 & 12.58 & 13.02 & 13.09& 13.10 \\
3& 13534& 9.004 & 12.27 & 13.34 & 13.20 & 13.26 & 13.31 & 13.52 & 13.54 & 13.61 \\
 6& 8105 & 9.009 & 14.51 & 15.11 & 14.80 & 14.57 & 14.36 & 14.29 & 14.17 & 14.19 \\
 8& 6242 & 9.011 & 15.65 & 16.33 & 15.81 & 15.47 & 15.13 & 14.80 & 14.60 & 14.55 \\
 10& 5479& 9.013 & 16.22 & 17.08 & 16.42 & 16.00 & 15.58 & 15.12 & 14.91 & 14.85 \\
\hline
\end{tabular}
\end{table*}

\vfill

\clearpage\eject

\begin{figure}

\centerline {\bf FIGURE CAPTIONS}
\vskip1cm

\caption[]{Cooling sequence for a 0.6 $\msol$ WD with the ion-electron contribution in the solid calculated by Potekhin \& Chabrier (2000) (solid line) or extrapolated from Yakovlev \& Shalybkov (1989) (dashed line). The arrow indicates the value $\Gamma_C=200$ for the plasma parameter at the center of the star.}
\label{fig1-cbfs-bw.eps}
\end{figure}

\begin{figure}
\caption[]{$L$-$T_C$ relations from our (solid line), Hansen (1999) (short-dash line), Wood (1995) (dotted-line) \& Montgomery et al. (1999) (long-dash line) calculations for a 0.6 $\msol$ WD with hydrogen and helium mass fractions $q(H)=10^{-4}$, $q(He)=10^{-2}$ and pure H-atmosphere.}
\label{fig2-cbfs-color.ps}
\end{figure}

\begin{figure}
\caption[]{Cooling sequences with (right curves) and without (left curves) crystallization-induced fractionation for a 0.6 $\msol$ DA WD.
Solid curves: calculations with our $L$-$T_C$ relation; long-dash curves: calculations with Hansen (1999) $L$-$T_C$ relation; dotted curves: calculations with Wood (1995) $L$-$T_C$ relation. Short-dash curve: Hansen (1999) cooling sequence. Dash-dot curve: present calculations with an initial profile obtained with a low $^{12}C(\alpha,\gamma)^{16}O$ reaction rate, with fractionation.}
\label{fig3-cbfs-color.ps}
\end{figure}

\begin{figure}
\caption[]{Cooling sequences $\mbol(t)$ for different masses for our
reference model DA WDs. The crosses indicate the cooling sequence for a 0.6 $\msol$ WD if the energy release due to C/O chemical fractionation at crystallization is not included in the cooling (see text).}
\label{fig4-cbfs-color.ps}
\end{figure}

\begin{figure}
\caption[]{Mass-$\te$ constant cooling times for H-atmosphere WDs. Ages are indicated in Gyr for each curve.}
\label{fig5-cbfs-bw.eps}
\end{figure}

\begin{figure}

\caption[]{Mass-$M_V$ constant cooling times for H-atmosphere WDs.}
\label{fig6-cbfs-bw.eps}
\end{figure}

\begin{figure}
\caption[]{$M_V$ vs. $(V-I)$ color-magnitude diagram for pure hydrogen atmosphere WDs. Constant WD cooling times correspond to the dotted lines. Cooling sequences for the 0.5, 0.8 and 1.2 $\msol$ are indicated by the dashed lines. The triangles correspond to the Leggett et al. (1998) WDs identified as H-rich atmosphere WDs.}
\label{fig7-cbfs-color.ps}
\end{figure}

\begin{figure}
\caption[]{Same as Figure 6 for a $M_J$ vs. $(J-K)$ color-magnitude diagram.}
\label{fig8-cbfs-color.ps}
\end{figure}

\begin{figure}
\caption[]{Color-color $(V-I)$-$(I-J)$ diagram for a 0.6 $\msol$ (solid line) and a 1.2 $\msol$ WD (dashed line). Cooling times are indicated in Gyr and are labelled for the 0.6 $\msol$ WD (solid dots). The squares correspond to the same ages for the 1.2 $\msol$ WD. For this mass the end of the track corresponds to 11.6 Gyr. The triangles correspond to the Leggett et al. (1998) WDs identified as H-rich atmosphere WDs.}
\label{fig9-cbfs-color.ps}
\end{figure}

\vfill\eject

\begin{figure}
\begin{center}
\epsfxsize=180mm
\epsfysize=180mm
\epsfbox{ 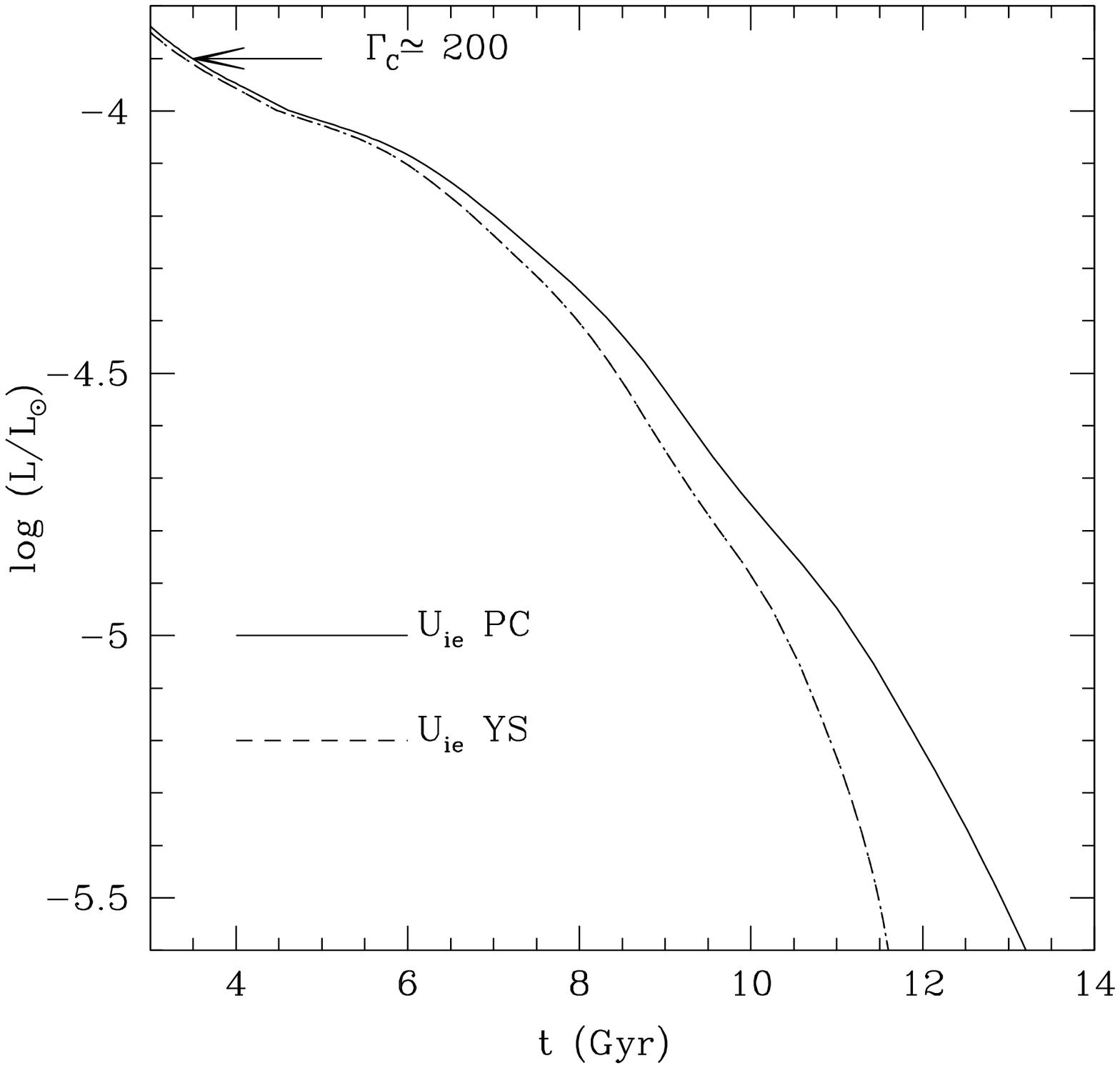}
\end{center}
\end{figure}

\vfill\eject

\begin{figure}
\begin{center}
\epsfxsize= 180mm
\epsfysize= 180mm
\epsfbox{ 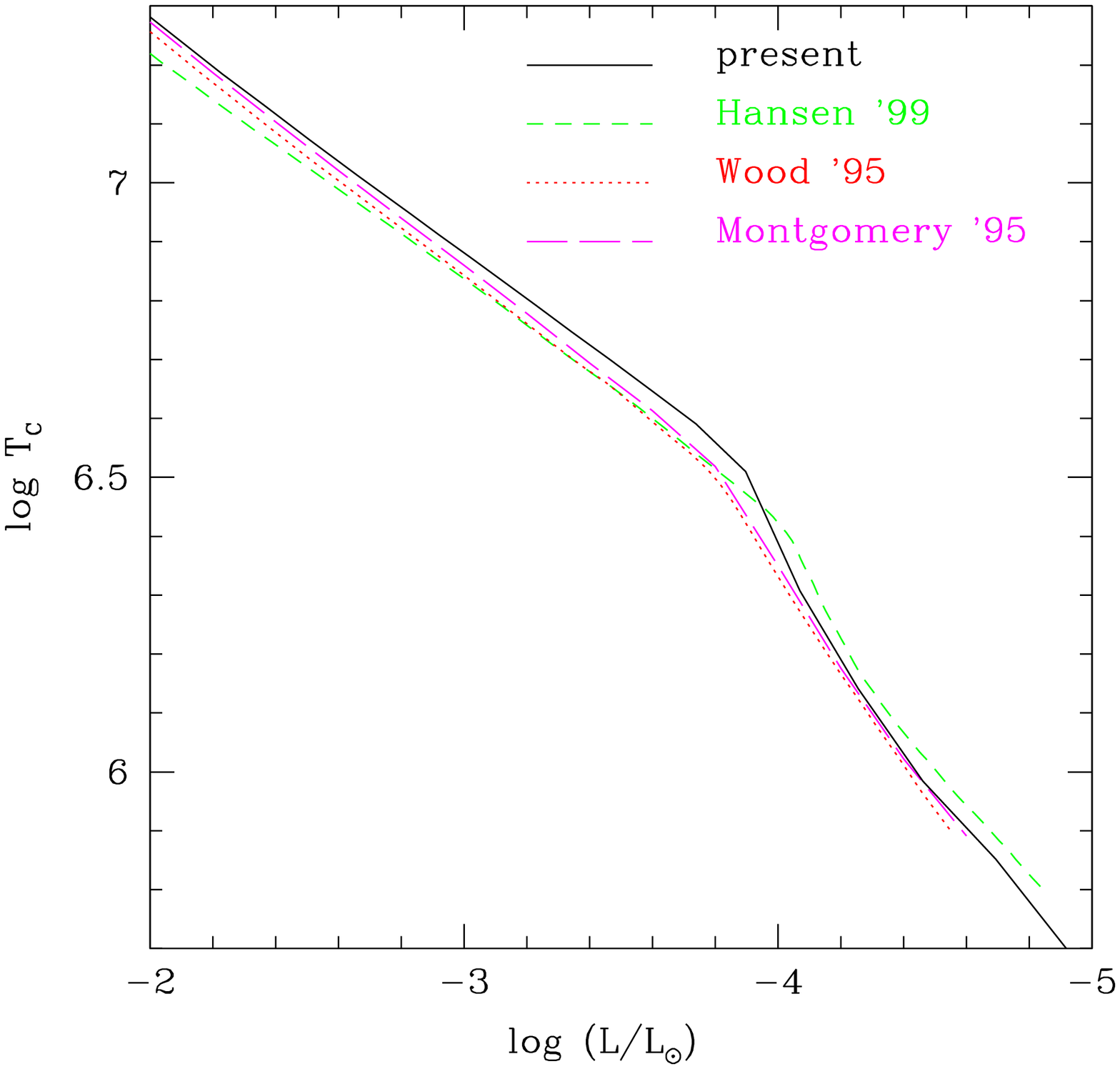}
\end{center}
\end{figure}

\vfill\eject

\begin{figure}
\begin{center}
\epsfxsize= 180mm
\epsfysize= 170mm
\epsfbox{ 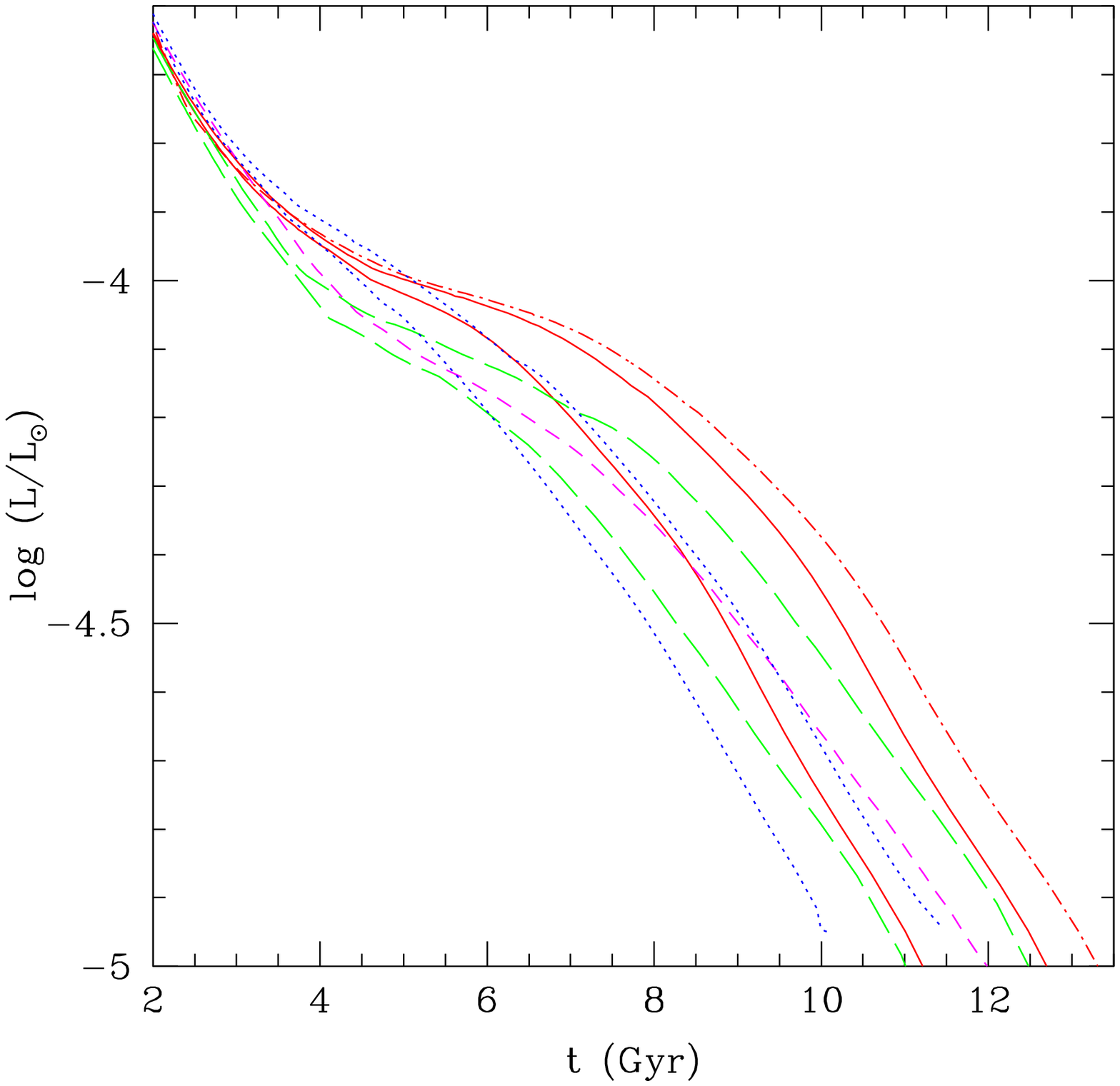}
\end{center}
\end{figure}

\vfill\eject

\begin{figure}
\begin{center}
\epsfxsize= 180mm
\epsfysize= 180mm
\epsfbox{ 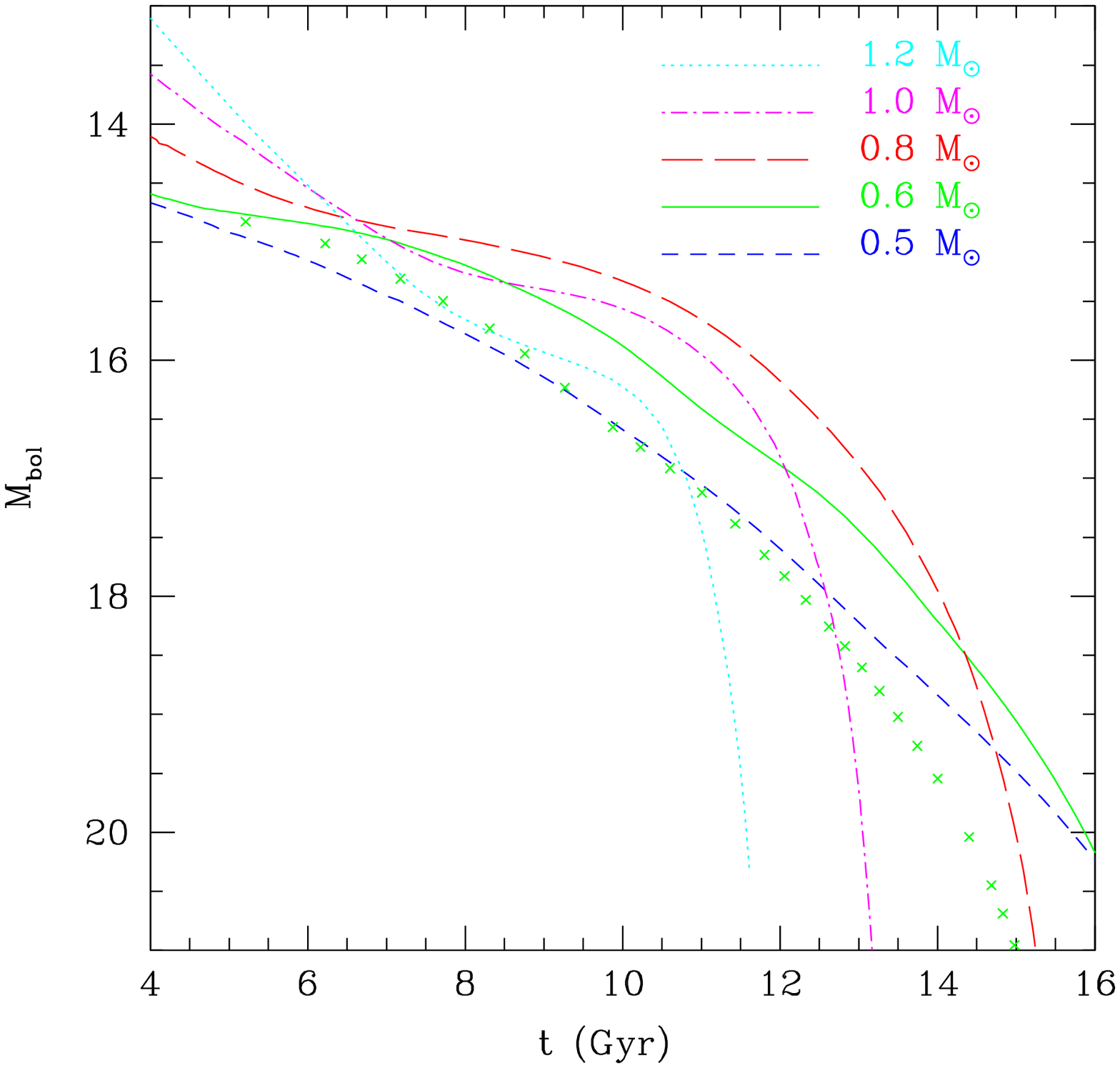}
\end{center}
\end{figure}

\vfill\eject

\begin{figure}
\begin{center}
\epsfxsize=180mm
\epsfysize=180mm
\epsfbox{ 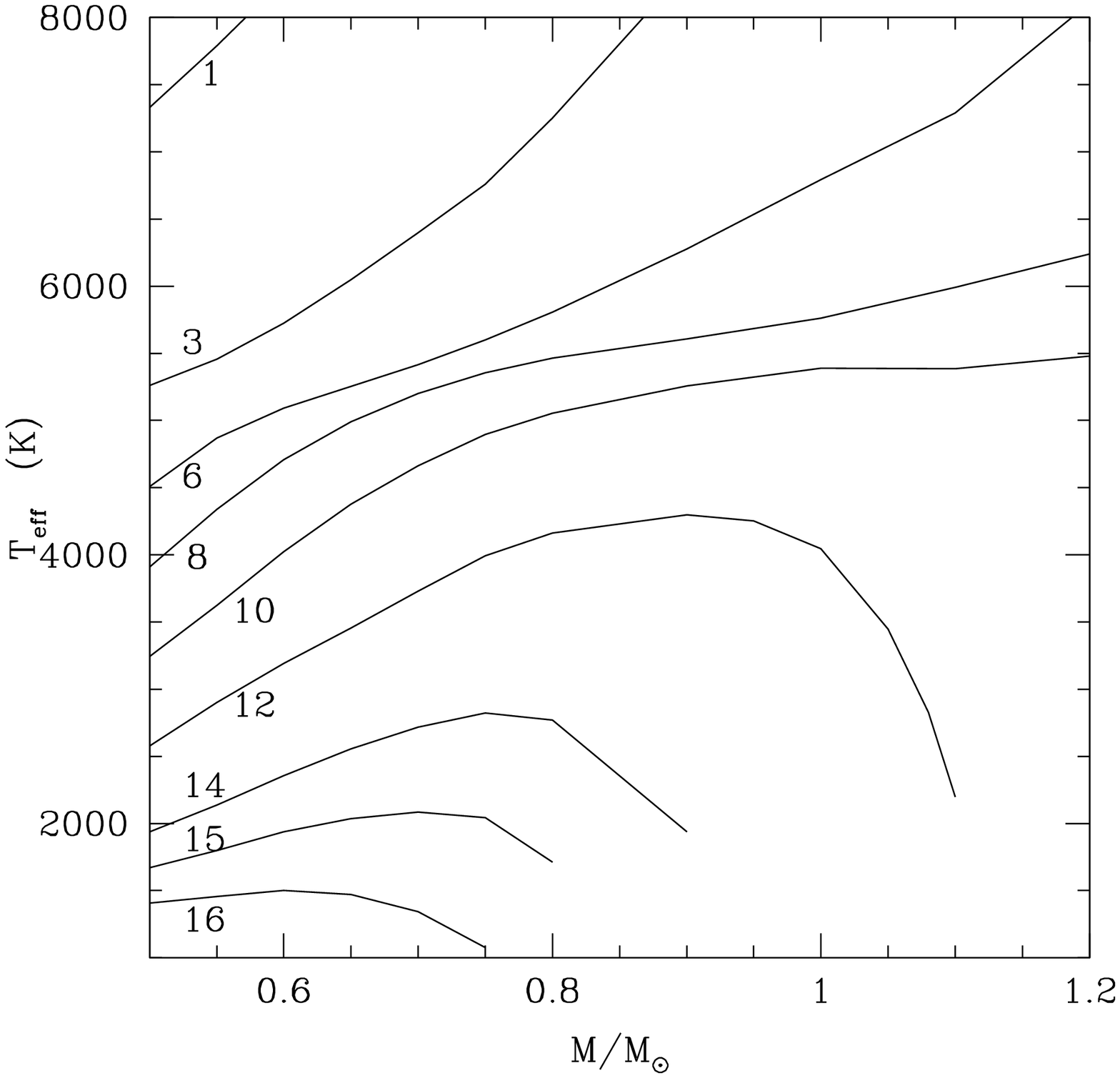}
\end{center}
\end{figure}

\vfill\eject

\begin{figure}
\begin{center}
\epsfxsize= 180mm
\epsfysize= 200mm
\epsfbox{ 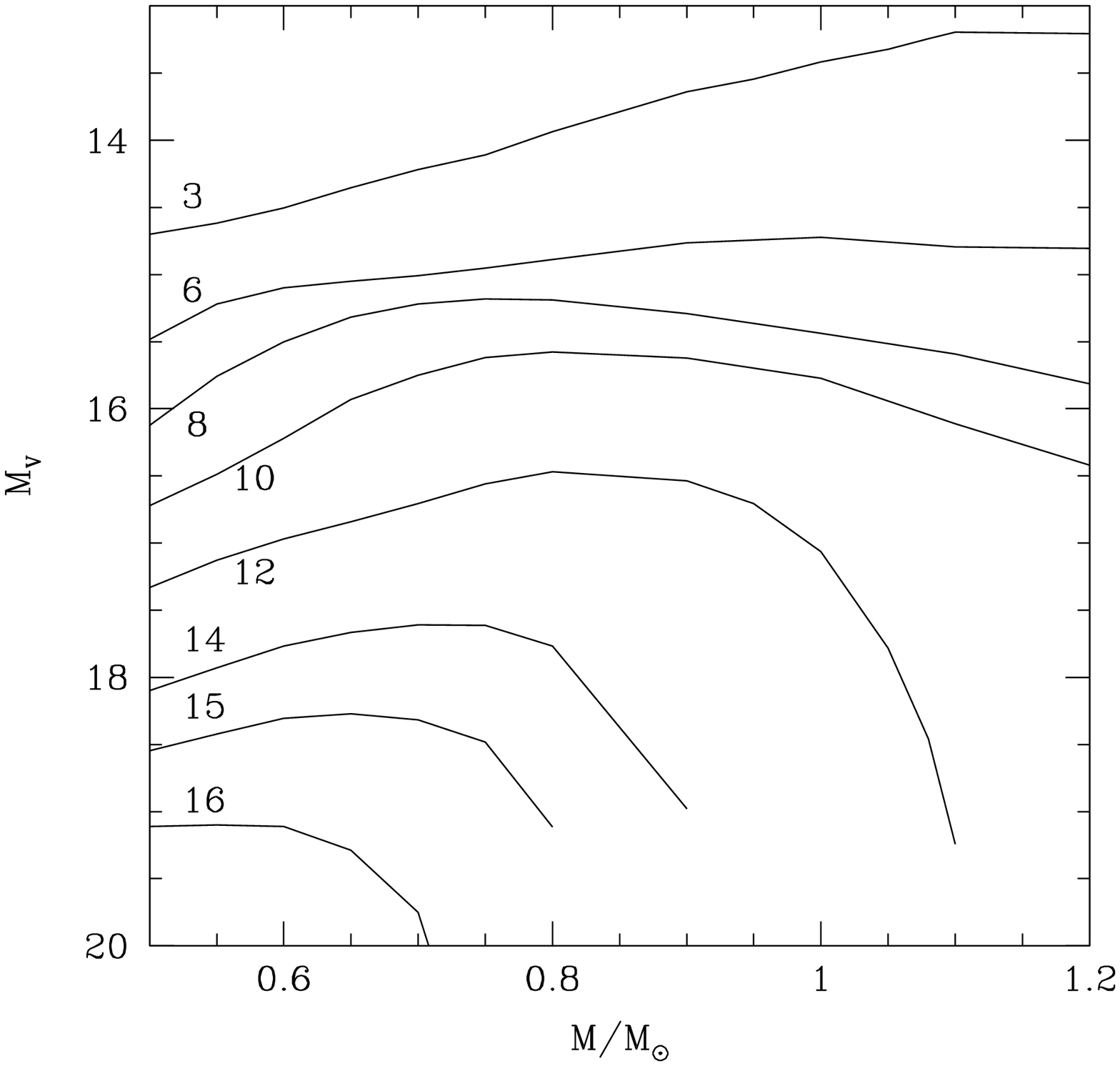}
\end{center}
\end{figure}

\vfill\eject

\begin{figure}
\begin{center}
\epsfxsize= 180mm
\epsfysize= 200mm
\epsfbox{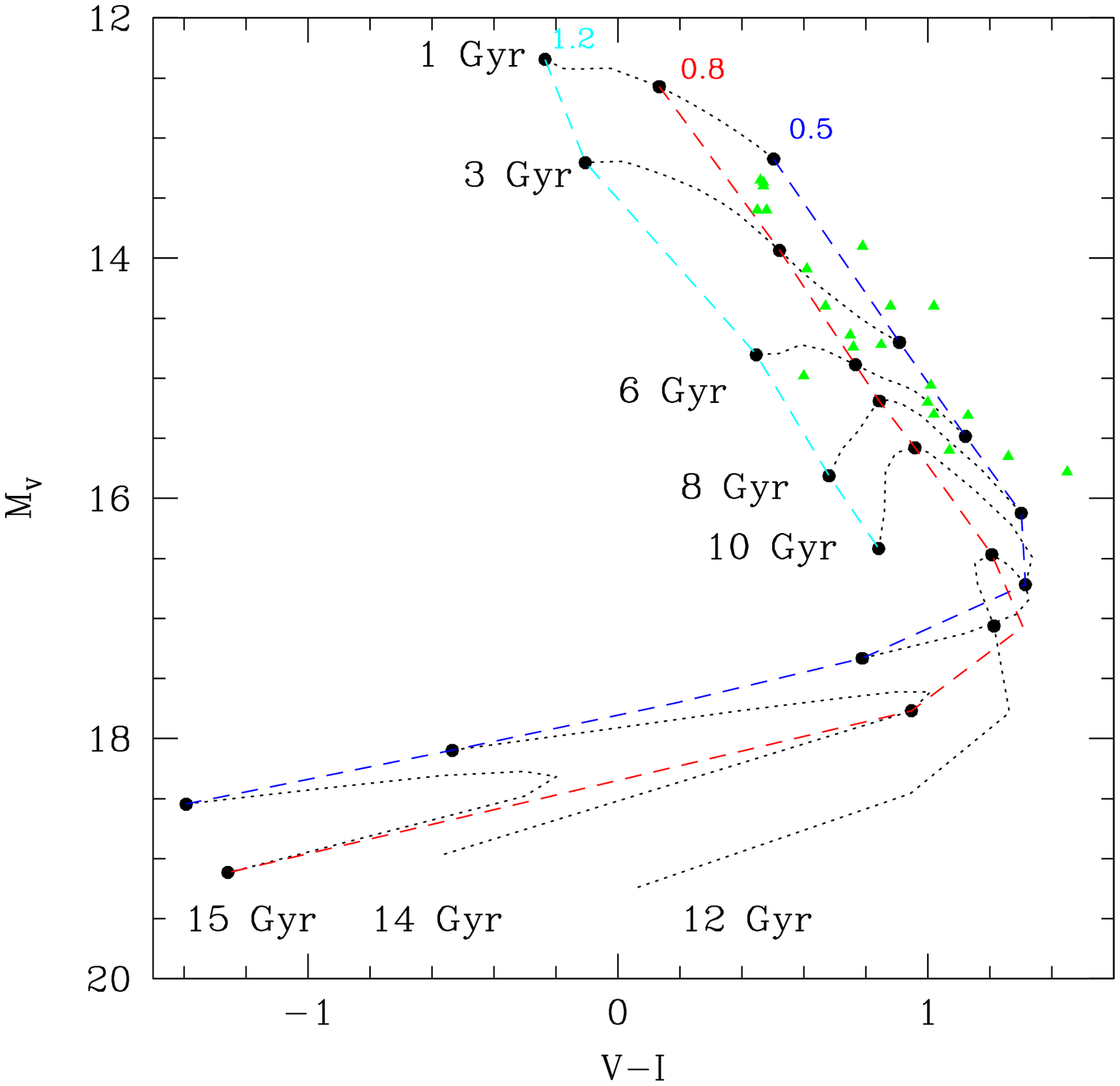}
\end{center}
\end{figure}

\vfill\eject

\begin{figure}
\begin{center}
\epsfxsize= 180mm
\epsfysize= 200mm
\epsfbox{ 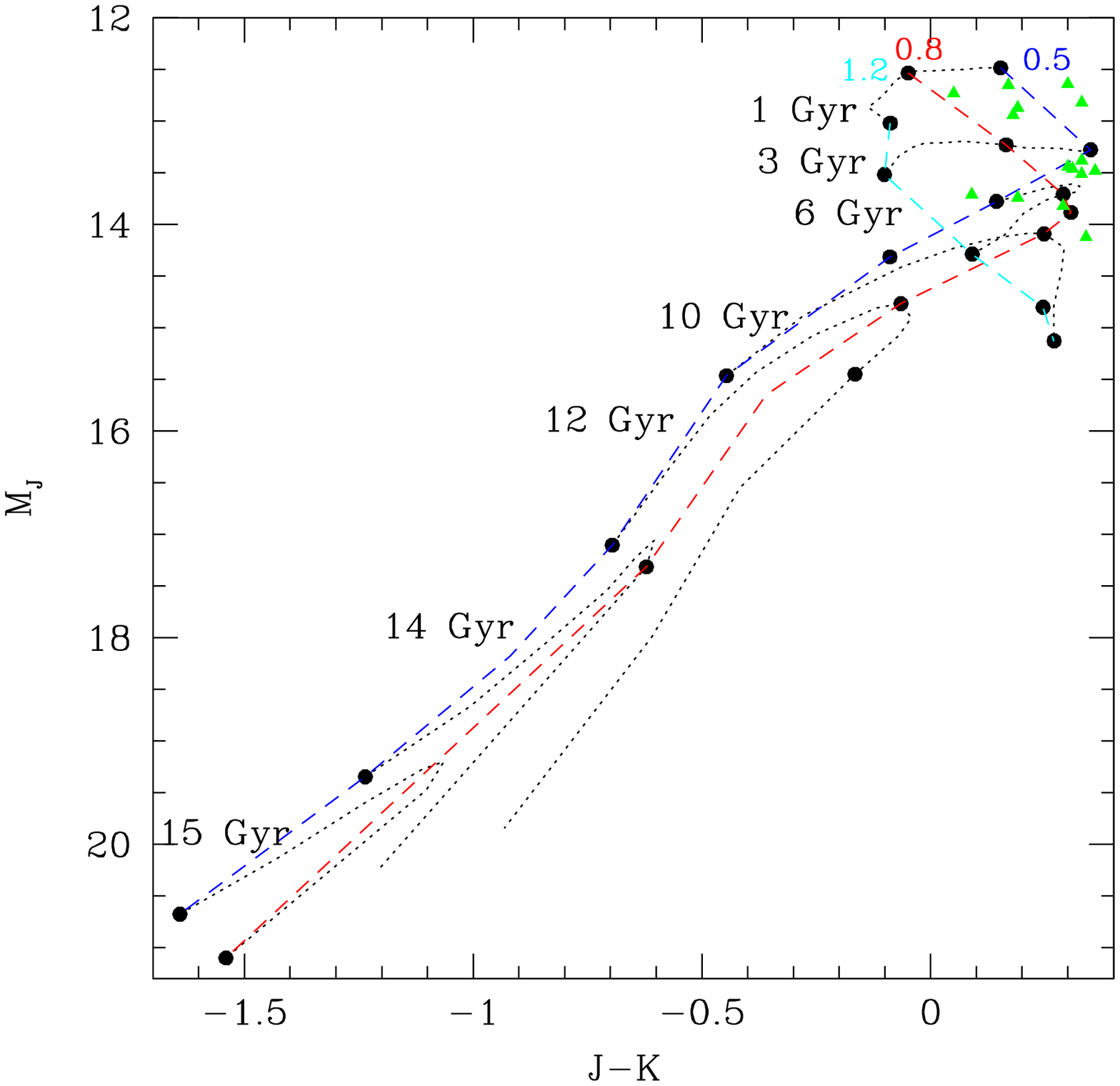}
\end{center}
\end{figure}

\vfill\eject

\begin{figure}
\begin{center}
\epsfxsize= 180mm
\epsfysize= 180mm
\epsfbox{ 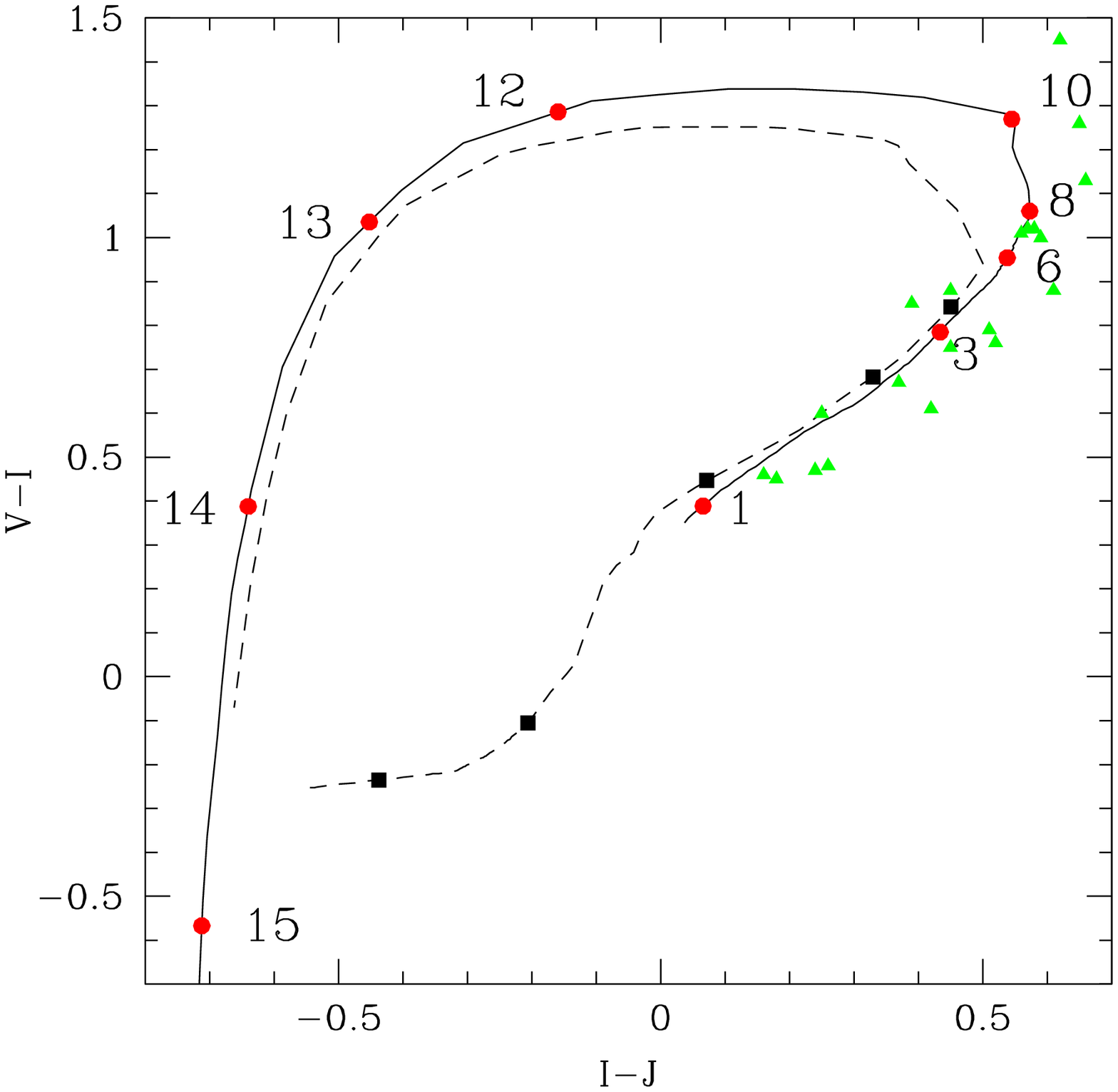}
\end{center}
\end{figure}

\vfill\eject

\end{document}